\documentclass[onecolumn,11pt,aps,nofootinbib,superscriptaddress,tightenlines]{revtex4}

\usepackage{amsmath}
\usepackage{amssymb}  
\usepackage{amsthm}
\usepackage{amsfonts}
\usepackage{graphicx}
\usepackage{bbm}
\usepackage{url}
\usepackage{dsfont}
\usepackage{color}
\newcommand{\tr}{\textnormal{tr}}

\newcommand{\cB}{\mathcal{B}}

\newcommand{\cH}{\mathcal{H}}

\newcommand{\cT}{\mathcal{T}}

\def\beq{\begin{equation}}
\def\eeq{\end{equation}}
\def\bq{\begin{quote}}
\def\eq{\end{quote}}
\def\ben{\begin{enumerate}}
\def\een{\end{enumerate}}
\def\bit{\begin{itemize}}
\def\eit{\end{itemize}}

\def\ra{\rightarrow}

\def\lb{\left(}
\def\rb{\right)}

\def\l|{\left|}
\def\r|{\right|}
\def\lbr{\left[}
\def\rbr{\right]}

\newcommand\C{\mathbb{C}}

\theoremstyle{plain}
\newtheorem{Theorem}{Theorem}

\theoremstyle{definition}

\newtheorem{innercustomthm}{Theorem}
\newenvironment{customthm}[1]
  {\renewcommand\theinnercustomthm{#1}\innercustomthm}
  {\endinnercustomthm}

\begin{document}
\title{Monotonicity of the Quantum Relative Entropy Under Positive Maps}
\author{Alexander M\"uller-Hermes}
\address{%
QMATH, Department of Mathematical Sciences, University of Copenhagen, 
2100 Copenhagen, Denmark}
\email{muellerh@posteo.net, muellerh@math.ku.dk}
\author{David Reeb}
\address{%
Institute for Theoretical Physics, Leibniz Universit\"at Hannover, 30167 Hannover, Germany}
\email{reeb.qit@gmail.com}

%\author{Alexander M\"uller-Hermes}
%\email{muellerh@posteo.net}
%\affiliation{Department of Mathematical Sciences, University of Copenhagen, 2100 Copenhagen, Denmark}
%%
%\author{David Reeb}
%\email{reeb.qit@gmail.com}
%\affiliation{Institute for Theoretical Physics, Leibniz Universit\"at Hannover, 30167 Hannover, Germany}

%
%
%
%

\begin{abstract}
We prove that the quantum relative entropy decreases monotonically under the application of any positive trace-preserving linear map, for underlying separable Hilbert spaces. This answers in the affirmative a natural question that has been open for a long time, as monotonicity had previously only been shown to hold under additional assumptions, such as complete positivity or Schwarz-positivity of the adjoint map. The first step in our proof is to show monotonicity of the sandwiched Renyi divergences under positive trace-preserving maps, extending a proof of the data processing inequality by Beigi [J.\ Math.\ Phys.\ \textbf{54}, 122202 (2013)] that is based on complex interpolation techniques. Our result calls into question several measures of non-Markovianity that have been proposed, as these would assess all positive trace-preserving time evolutions as Markovian.
\end{abstract}

\maketitle

\section{Introduction}

For any pair of quantum states $\rho,\sigma$ acting on the same Hilbert space $\cH$, the \emph{quantum relative entropy} is defined by
\begin{equation}\label{defineD}
 D(\rho\|\sigma):=
\begin{cases} 
\tr[\rho(\log\rho-\log\sigma)], & \mbox{if }  \mbox{supp}[\rho]\subseteq\mbox{supp}[\sigma]
\\ +\infty, & \mbox{otherwise}.
\end{cases}
\end{equation}
It was first introduced by Umegaki \cite{umegaki1962} as a quantum generalization of the classical Kullback-Leibler divergence between probability distributions \cite{kullbackleibler}. Both are distance-like measures, also called \emph{divergences}, and have found a multitude of applications and operational interpretations in diverse fields like information theory, statistics, and thermodynamics. We refer to the review \cite{wehrlreview} and the monograph \cite{ohyapetz} for more details on the quantum relative entropy.

The key property of divergence measures in applications is that they should not increase under any evolution $\Phi:\tau\mapsto\Phi(\tau)$ that is allowed in a certain context. This means that a divergence $D$ should, for any admissible mapping $\Phi$, be \emph{monotonic} in the sense that
\begin{equation}
D(\rho\| \sigma)\geq D(\Phi(\rho)\|\Phi(\sigma))\qquad\forall\,\rho,\sigma.
\label{equ:DataProc}
\end{equation}

For the quantum relative entropy $D$ defined in (\ref{defineD}), the monotonicity property (\ref{equ:DataProc}) has indeed been shown to hold for several classes of maps $\Phi$. In \cite{lindblad-cmp1975} it was proven for all \emph{completely positive} trace-preserving linear maps $\Phi:\cT(\cH)\to\cT(\cH')$ between trace-class operators, which model the most general quantum mechanical time-evolutions in the presence of entanglement \cite{lindblad-cmp1975,nielsenchuang}. The proof relied on the Stinespring dilation theorem \cite{stinespring1955} and strong subadditivity of the von Neumann entropy \cite{lieb-ruskai-ssa-jmp}; cf.\ also \cite{ruskai-review}. A different proof for the monotonicity was given in \cite{uhlmann-cmp1977} under the strictly weaker condition that $\Phi:\cT(\cH)\to\cT(\cH')$ be a linear trace-preserving map whose adjoint $\Phi^*:\cB(\cH')\to\cB(\cH)$ is a \emph{Schwarz map}, i.e.\ satisfies $\Phi^*(X^\dagger X)\geq \Phi^*(X)^\dagger \Phi^*(X)$ for all $X\in\cB(\cH')$; cf.\ also \cite{ohyapetz,petz-2003}. The class of maps for which monotonicity is known has been further extended to contain the transposition map along with compositions and convex combinations of monotonic maps \cite{hiai2011}. Via the operational interpretation of the quantum relative entropy in binary hypothesis testing, monotonicity has also been proven in \cite[Theorem 5.5]{hayashi-book} for all trace-preserving \emph{tensor-stable positive maps} \cite{ts-positive-paper}, although it is not known whether this includes any map beyond compositions of a completely positive map with either the identity or the transposition map.

The question of monotonicity of the quantum relative entropy has however remained open \cite{hiai2011,ohyapetz,petz-2003,ruskai-review} for the natural class of linear maps $\Phi:\cT(\cH)\to\cT(\cH')$ that are subject only to the constraint that they map quantum states to quantum states. These are exactly the \emph{positive} trace-preserving linear maps $\Phi$. This class strictly contains all previously known ones.

Here we show that the quantum relative entropy is indeed monotonically decreasing under any positive trace-preserving linear map between trace-class operators, for separable underlying Hilbert spaces. To show our result we first observe that Beigi's proof \cite{beigi} of the monotonicity of certain \emph{sandwiched Renyi divergences} \cite{muller2013quantum,winter-wilde-yang} extends to general positive trace-preserving maps, and after a limiting argument we lift our statement to infinite dimensions.

Besides generalizing all previously known cases of monotonicity in the most natural way, our result is also relevant when considering quantum-mechanical time evolutions of full systems only, as e.g.\ in \cite{weinberg-qm-vol2}, without ``innocent bystanders''. In such a framework, the physical time evolutions are the positive trace-preserving maps. Our result then asserts that the data processing inequality in (\ref{equ:DataProc}) -- an operationally very reasonable statement -- does indeed hold for all physically allowed evolutions.

On the other hand, the result shows that some proposed measures to assess the (non\nobreakdash-)Markovianity of quantum time evolutions \cite{ buscemiprl,non-M-review} have deficiencies beyond those already known. In particular, measures based on the quantum relative entropy  \cite{LPB,non-M-review,ushadevi} cannot detect non-Markovian behaviour in any time evolution that is merely positive. Since the commonly accepted notion of Markovianity is however that of completely positive transition maps \cite{non-M-review,wolf-cirac}, our results enlarge the set of time evolutions for which the measures from \cite{LPB,ushadevi} are known to be insufficient.

\bigskip
\textbf{Notation.} For a complex separable Hilbert space $\cH$ we denote the set of trace-class operators by $\cT(\cH)$ and its positive semidefinite elements by $\cT_+(\cH)$. We consider the relative entropy $D(\rho\|\sigma)$ by a literal extension of definition (\ref{defineD}) for general positive semidefinite trace-class operators $\rho,\sigma$ \cite{lindblad-cmp1974}, not only for density operators which are of trace $1$. A linear map $\Phi:\cT(\cH)\to\cT(\cH')$ is called \emph{trace-nonincreasing} if ${\rm tr}[\Phi(A)]\leq{\rm tr}[A]$ for all $A\in\cT_+(\cH)$, and \emph{trace-preserving} if the same relation holds with equality; these two properties are equivalent to $\Phi^*({\mathbbm 1}_{\cH'})\leq{\mathbbm 1}_{\cH}$ resp.\ $\Phi^*({\mathbbm 1}_{\cH'})={\mathbbm 1}_{\cH}$, where $\Phi^*:\cB(\cH')\to\cB(\cH)$ denotes the adjoint mapping. A linear map $\Phi:\cT(\cH)\to\cT(\cH')$ is called \emph{positive} if $\Phi(A)\geq0$ for all $A\in\cT_+(\cH)$, and \emph{completely positive} if $(\Phi\otimes{\rm id}_n):\cT(\cH\otimes\C^n)\to\cT(\cH'\otimes\C^n)$ is positive for all $n\in{\mathbb N}$.

\section{Results}

\begin{Theorem}[Monotonicity of the quantum relative entropy under positive trace-preserving maps]
Let $\Phi:\cT(\cH)\to\cT(\cH')$ be a positive trace-preserving linear map, where $\cH$ and $\cH'$ are separable Hilbert spaces. Then, for any positive semidefinite operators $\rho,\sigma\in\cT_+(\cH)$ we have
\begin{equation}
D(\rho\|\sigma)\geq D(\Phi(\rho)\|\Phi(\sigma)).
\label{equ:DataProcPos}
\end{equation}
\label{thm:Main}
\end{Theorem}

The proof of this statement makes use of the family of \emph{sandwiched Renyi-$\alpha$ divergences} $D_\alpha$ which were recently introduced in \cite{muller2013quantum,winter-wilde-yang}. For parameter $\alpha\in(1,\infty)$ and positive semidefinite operators $\rho,\sigma\in\cT_+(\C^d)$ on a finite-dimensional space $\C^d$ they are defined as
\begin{equation}
 D_\alpha\lb\rho||\sigma\rb:=
\begin{cases} 
\frac{1}{\alpha-1}\log\lb\tr\left[\lb\sigma^{\frac{1-\alpha}{2\alpha}}\rho\sigma^{\frac{1-\alpha}{2\alpha}}\rb^\alpha\right]\rb, & \mbox{if }  
\text{supp}[\rho]\subseteq\text{supp}[\sigma]
\\ +\infty, & \mbox{otherwise}.
\end{cases}
\label{equ:Renyi}
\end{equation}  
The monotonicity property for these divergence measures, i.e.\ the analogue of Eq.\ (\ref{equ:DataProc}), under completely positive trace-preserving linear maps $\Phi$ has been established for various ranges of the Renyi parameter $\alpha$ in \cite{beigi,frank-lieb,muller2013quantum,winter-wilde-yang}. We show that the proof in \cite{beigi} also works under the weaker requirement of positivity instead of complete positivity (and also when requiring the map to be merely trace-nonincreasing rather than trace-preserving):

\begin{Theorem}[Monotonicity of sandwiched Renyi divergences under positive trace-nonincreasing maps]
Let $\Phi:\cT(\C^d)\ra\cT(\C^{d'})$ be a positive trace-nonincreasing linear map and $\alpha\in(1,\infty)$. Then, for any positive semidefinite operators $\rho,\sigma\in\cT_+(\C^d)$ we have
\begin{equation}
D_\alpha(\rho\|\sigma)\geq D_\alpha(\Phi(\rho)\| \Phi(\sigma)).
\label{equ:DataProcPosAlpha}
\end{equation}
\label{thm:DataProcSand}
\end{Theorem}

Whereas the requirement of a trace-nonincreasing positive $\Phi$ is enough to show the monotonicity of $D_{\alpha}$ (Theorem \ref{thm:DataProcSand}), this weakening does not carry over to the monotonicity of the quantum relative entropy $D$ (Theorem \ref{thm:Main}) where we assumed a trace-preserving $\Phi$, due to the normalization required of $\rho$ for the convergence $\lim_{\alpha\to1}D_\alpha(\rho\|\sigma)=D(\rho\|\sigma)$ \cite{muller2013quantum,winter-wilde-yang} in Step 1 of the proof of Theorem \ref{thm:Main}. One can in fact easily convince oneself that the relative entropy is not generally monotonic under positive trace-nonincreasing linear maps, even between finite-dimensional states and in the completely positive case; an (essentially classical, i.e.\ commutative) counterexample is the map $\Phi:\cT(\C^2)\to\cT(\C^2)$ defined as $\Phi\big(\left(\begin{smallmatrix}v&w\\x&y\end{smallmatrix}\right)\big):=\left(\begin{smallmatrix}v/2&0\\0&y\end{smallmatrix}\right)$ along with the states $\rho:=\left(\begin{smallmatrix}1/3&0\\0&2/3\end{smallmatrix}\right)$, $\sigma:=\left(\begin{smallmatrix}2/3&0\\0&1/3\end{smallmatrix}\right)$, which gives $D(\rho\|\sigma)=(\log2)/3<(\log2)/2=D(\Phi(\rho)\|\Phi(\sigma))$. However, the proof of Theorem \ref{thm:Main} can be extended to yield monotonicity of $D$ under trace-nonincreasing positive maps that obey a state-dependent form of normalization:
\begin{customthm}{1'}[Generalization of Theorem \ref{thm:Main} to trace-nonincreasing maps]
\textit{For separable Hilbert spaces $\cH$ and $\cH'$, let $\Phi:\cT(\cH)\to\cT(\cH')$ be a positive trace-nonincreasing linear map and $\rho,\sigma\in\cT_+(\cH)$ be operators satisfying ${\rm tr}[\Phi(\rho)]={\rm tr}[\rho]$. Then we have}
\begin{equation}
D(\rho\|\sigma)\geq D(\Phi(\rho)\|\Phi(\sigma)).
\end{equation}
\label{theroremMainPrime}
\end{customthm}The presuppositions of Theorem \ref{theroremMainPrime} actually imply, by similar steps as in its proof, that $\Phi$ is a trace-\emph{preserving} positive linear map when restricted to the support of $\rho$. But this subspace may be strictly smaller than ${\rm supp}[\sigma]\subseteq\cH$ even in the nontrivial case $D(\rho\|\sigma)<\infty$. Therefore in particular, Theorem \ref{theroremMainPrime} generalizes monotonicity results for trace-nonincreasing maps from \cite{hiai2011}.

Finally, note that Theorems \ref{thm:Main} and \ref{theroremMainPrime} show monotonicity of the quantum relative entropy $D$ for any separable Hilbert spaces (finite- or infinite-dimensional), whereas we defined the sandwiched Renyi divergences $D_\alpha$ occurring in Theorem \ref{thm:DataProcSand} only for finite-dimensional underlying spaces. Two definitions of sandwiched Renyi-$\alpha$ divergences have been suggested for infinite-dimensional von Neumann algebras in recent preprints \cite{volkherVNalgebras,jencovaVNalgebras}. One of these quantities was proven to be monotonic under positive trace-preserving maps \cite{jencovaVNalgebras} by extending our proof, whereas \cite{volkherVNalgebras} needed complete positivity to show monotonicity. The monotonicity of the quantum relative entropy under positive maps, i.e.\ our main Theorem \ref{thm:Main} (or \ref{theroremMainPrime}), was however not obtained in these works, since it remains open whether the quantities defined in \cite{jencovaVNalgebras} converge to the quantum relative entropy $D$ in the limit $\alpha\to1$.

\section{Proofs}

We first prove Theorem \ref{thm:DataProcSand}, recapitulating the steps of Beigi's proof \cite{beigi} to make an additional observation that gives monotonicity of $D_\alpha$ for all positive trace-nonincreasing maps. The proofs of Theorems \ref{thm:Main} and \ref{theroremMainPrime}, which cover also general separable Hilbert spaces, follow afterwards.

\bigskip
\noindent\textbf{Proof of Theorem \ref{thm:DataProcSand}.} Note that $\text{supp}[\rho]\subseteq\text{supp}[\sigma]$ implies $\text{supp}[\Phi(\rho)]\subseteq\text{supp}[\Phi(\sigma)]$ due to positivity of $\Phi$, whereas there is nothing to prove in the case $\text{supp}[\rho]\not\subseteq\text{supp}[\sigma]$. We can thus restrict attention on both sides of (\ref{equ:DataProcPosAlpha}) to the supports of $\sigma$ and $\Phi(\sigma)$, respectively, and w.l.o.g.\ assume that both $\sigma$ and $\Phi(\sigma)$ are of full rank.

The sandwiched Renyi divergences can be written in terms of non-commutative $L_p$-norms $\|\cdot\|_{p,\sigma}$ on $\cT(\C^d)$ for $p\in[1,\infty)$, defined for $X\in\cT(\C^d)$ by
\[
\| X\|_{p,\sigma} := \tr\lbr\left|\sigma^{\frac{1}{2p}}X\sigma^{\frac{1}{2p}}\right|^p\rbr^{\frac{1}{p}};
\]
for $p=\infty$ we define $\|X\|_{\infty,\sigma}:=\lim_{p\to\infty}\| X\|_{p,\sigma}=\|X\|_\infty$. With the map $\Gamma_\sigma:\cT(\C^d)\ra\cT(\C^d)$ defined by $\Gamma_\sigma(X): = \sigma^{1/2}X\sigma^{1/2}$, one can then write \cite{beigi}:
\[
D_\alpha\lb\rho\|\sigma\rb = \frac{1}{\alpha-1}\log\lb\| \Gamma_\sigma^{-1}\lb\rho\rb\|_{\alpha,\sigma}^\alpha\rb.
\]
For any linear map $\Psi:\cT(\C^d)\to\cT(\C^{d'})$ we furthermore define the \emph{induced operator norm} w.r.t.\ the norms $\|\cdot\|_{p,\sigma}$ and $\|\cdot\|_{p,\Phi(\sigma)}$ on the in- and output spaces, respectively:
\[
\|\Psi\|_{(p,\sigma)\ra(p,\Phi(\sigma))}:=\sup_{X\in\cT(\C^d)\setminus\{0\}}\frac{\|\Psi(X)\|_{p,\Phi(\sigma)}}{\|X\|_{p,\sigma}}.
\]

We now consider $X:=\Gamma_{\sigma}^{-1}(\rho)=\sigma^{-1/2}\rho\sigma^{-1/2}$, so that the desired inequality (\ref{equ:DataProcPosAlpha}) becomes equivalent to $\|X\|_{\alpha,\sigma}\geq\|\Gamma^{-1}_{\Phi(\sigma)}\circ\Phi\circ\Gamma_\sigma(X)\|_{\alpha,\Phi(\sigma)}$. In order to show the theorem, it is therefore sufficient to prove the following bound on the induced operator norm:
\begin{equation}
\|\Gamma^{-1}_{\Phi(\sigma)}\circ \Phi\circ \Gamma_\sigma\|_{(\alpha,\sigma)\ra (\alpha,\Phi(\sigma))} \leq 1.
\label{equ:NormContr}
\end{equation}

To show \eqref{equ:NormContr}, Beigi \cite{beigi} applies the following Riesz-Thorin-type interpolation result:
\begin{Theorem}[Riesz-Thorin theorem for $L_{p,\sigma}$ spaces \cite{beigi,berghloefstroem}]
Let $\Psi:\cT(\C^d)\ra\cT(\C^{d'})$ be a linear map, $1\leq p_0\leq p_1\leq \infty$, and define $p_\theta$ for $\theta\in[0,1]$ via
\[
\frac{1}{p_\theta} = \frac{1-\theta}{p_0} + \frac{\theta}{p_1}.
\]
Then we have, for any operators $\sigma\in\cT(\C^d)$ and $\sigma'\in\cT(\C^{d'})$ that are strictly positive definite:
\begin{equation}\label{riesz-thorin-eqn}
\| \Psi\|_{(p_\theta,\sigma)\ra (p_\theta,\sigma')} \leq \| \Psi\|^{1-\theta}_{(p_0,\sigma)\ra (p_0,\sigma')} \| \Psi\|^{\theta}_{(p_1,\sigma)\ra (p_1,\sigma')} .
\end{equation}
\label{thm:RieszT}
\end{Theorem}

Setting $p_0 = 1$, $p_1=\infty$, $\sigma'=\Phi(\sigma)$, and $\Psi = \Gamma^{-1}_{\Phi(\sigma)}\circ \Phi\circ \Gamma_\sigma$ in Theorem \ref{thm:RieszT}, one can evaluate the operator norms on the right-hand side of (\ref{riesz-thorin-eqn}):
\begin{align*}
\|\Gamma^{-1}_{\Phi(\sigma)}\circ \Phi \circ \Gamma_\sigma\|_{(1,\sigma)\ra (1,\Phi(\sigma))} &= \sup_{X\in\cT(\C^d)}\frac{\|\Phi\circ\Gamma_\sigma(X)\|_1}{\|\Gamma_\sigma(X)\|_1} \\
&= \sup_{Y\in\cT(\C^d)}\frac{\|\Phi(Y)\|_1}{\|Y\|_1}=:\| \Phi\|_{1\ra 1},
\end{align*}
where $\|\cdot\|_1$ denotes the usual trace norms on $\cT(\C^d)$ and $\cT(\C^{d'})$, respectively, and $\|\cdot\|_{1\to1}$ is the operator norm induced by these.

At this point, Beigi \cite{beigi} assumed $\Phi$ to be completely positive and trace-preserving in order to conclude $\|\Phi\|_{1\to1}=1$. However, $\|\Phi\|_{1\to1}\leq1$ holds already for any positive and trace-nonincreasing linear map $\Phi$, as can be seen from the Russo-Dye Theorem \cite[Corollary 2.9]{paulsen2002} applied to the adjoint map $\Phi^*$. This is a positive map and satisfies $0\leq\Phi^*(\mathbbm{1})\leq\mathbbm{1}$ due to the trace-nonincreasing property of $\Phi$, so that the Russo-Dye Theorem yields $\| \Phi\|_{1\ra 1} = \| \Phi^*\|_{\infty\ra\infty}=\|\Phi^*(\mathbbm{1})\|_\infty\leq\|\mathbbm{1}\|_\infty=1$. We can therefore conclude $\|\Gamma^{-1}_{\Phi(\sigma)}\circ \Phi\circ \Gamma_\sigma\|_{(1,\sigma)\ra (1,\Phi(\sigma))}\leq1$.

The second term in (\ref{riesz-thorin-eqn}) can be evaluated with the Russo-Dye Theorem as well:
\begin{align*}
\|\Gamma^{-1}_{\Phi(\sigma)}\circ \Phi\circ \Gamma_\sigma\|_{(\infty,\sigma)\ra (\infty,\Phi(\sigma))} &=\|\Gamma^{-1}_{\Phi(\sigma)}\circ \Phi\circ \Gamma_\sigma\|_{\infty\ra\infty}\\
 &= \| \Gamma^{-1}_{\Phi(\sigma)}\circ \Phi\circ \Gamma_\sigma\lb\mathbbm{1}\rb\|_{\infty} = \| \mathbbm{1}\|_\infty = 1.
\end{align*}

Choosing $\theta = (\alpha-1)/\alpha$ gives $p_\theta=\alpha$, so that the inequality (\ref{equ:NormContr}) follows from Theorem \ref{thm:RieszT}.\qed % end of first proof (Theorem 2)

\bigskip
\bigskip

\noindent\textbf{Proof of Theorem \ref{thm:Main}.} We split the proof into two steps.

\smallskip

\noindent\textit{Step 1: Proof for finite-dimensional Hilbert spaces.} The statement of Theorem \ref{thm:Main} for finite-dimensional $\cH=\C^d$, $\cH'=\C^{d'}$ follows quickly from Theorem \ref{thm:DataProcSand} by taking the limit $\alpha\to1$ on both sides of the inequality and noting that $\lim_{\alpha\ra 1}D_\alpha(\rho\|\sigma) = D(\rho\|\sigma)$ for $\tr[\rho]=1$, as shown in \cite{muller2013quantum,winter-wilde-yang}. For $\tr[\rho]>0$, Theorem \ref{thm:Main} follows from this by using $D(\rho\|\sigma)=\tr[\rho]\left(D(\rho/\tr[\rho]\|\sigma)+\log\tr[\rho]\right)$ along with $\tr[\Phi(\rho)]=\tr[\rho]$, whereas $\tr[\rho]=0$ implies $\rho=0$ and both sides of the inequality (\ref{equ:DataProcPos}) vanish.

\medskip

\noindent\textit{Step 2: Extension to infinite dimensions.} Let $\cH$ and $\cH'$ now be general separable Hilbert spaces (not necessarily finite-dimensional). We lift our finite-dimensional result from Step 1 to this case, by applying a similar technique as \cite{lindblad-cmp1975,shirokovholevoconcept} to general positive maps. To do this, let $P_n\in\cB(\cH)$ be a sequence of finite-dimensional orthogonal projections such that $P_n\to{\mathbbm 1}_\cH$ converges strongly as $n\to\infty$ (i.e., $\lim_{n\to\infty}\|(P_n-{\mathbbm1}_\cH)\psi\|=0$ for all $\psi\in\cH$); similarly, let $P'_n\in\cB(\cH')$ be a sequence of finite-dimensional projections with $P'_n\to{\mathbbm 1}_{\cH'}$ strongly. Define now the maps $\Phi_n:\cT(\cH)\to\cT(\cH')$ as
\begin{equation}
\Phi_n(A):=P'_n\Phi(P_nAP_n)P'_n+\frac{P'_n}{{\rm tr}[P'_n]}{\rm tr}[\Phi(P_nAP_n)({\mathbbm 1}_\cH-P'_n)].\label{definePhin}
\end{equation}

We first show that for all $A\in\cT(\cH)$, $\Phi_n(A)$ converges in trace norm to $\Phi(A)$; this is termed \emph{strong convergence} of the maps $\Phi_n$ to $\Phi$ \cite{shirokovholevoconcept}. Similar convergence results are shown in \cite{lindblad-cmp1975,shirokovholevoconcept}, and applied in related contexts for example in \cite{holevoshirokovapply}; since these prior results were stated for completely positive maps $\Phi$ and for somewhat different approximations $\Phi_n$, we give some detail on the argument. Observe, therefore, that the triangle inequality yields
\begin{align*}
\|\Phi(A)-\Phi_n(A)\|_1\leq\,\,&\|\Phi(A)-P'_n\Phi(A)P'_n\|_1 +\|P'_n\Phi(A-P_nAP_n)P'_n\|_1+\cdots \\ &\cdots+|{\rm tr}[\Phi(P_nAP_n)({\mathbbm1}_{\cH'}-P'_n)]|.
\end{align*}
Here, the first term $\|\Phi(A)-P'_n\Phi(A)P'_n\|_1\leq\|({\mathbbm1}-P'_n)\Phi(A)\|_1+\|P'_n\Phi(A)({\mathbbm1}-P'_n)\|_1\leq\|({\mathbbm1}-P'_n)\Phi(A)\|_1+\|\Phi(A)({\mathbbm1}-P'_n)\|_1$ converges to $0$ as $({\mathbbm1}-P'_n)\to0$ strongly and $\Phi(A)\in\cT(\cH')$. The second term converges to $0$ due to $\|P'_n\Phi(A-P_nAP_n)P'_n\|_1\leq\|A-P_nAP_n\|_1$, using that $\Phi$ is positive and trace-preserving and by similar reasoning as before. The third term $|{\rm tr}[\Phi(P_nAP_n-A)({\mathbbm1}-P'_n)+\Phi(A)({\mathbbm1}-P'_n)]|\leq\|P_nAP_n-A\|_1\|{\mathbbm1}-P'_n\|_\infty+\|\Phi(A)({\mathbbm1}-P'_n)\|_1$ vanishes for $n\to\infty$ as well. As the quantum relative entropy is lower-semicontinuous \cite{holevobook,lindblad-cmp1975,ohyapetz,wehrlreview}, the trace-norm convergence statements $\Phi_n(\rho)\to\Phi(\rho)$ and $\Phi_n(\sigma)\to\Phi(\sigma)$ give
\begin{align}
D(\Phi(\rho)\|\Phi(\sigma))&\leq\liminf_{n\to\infty}D(\Phi_n(\rho)\|\Phi_n(\sigma))\nonumber\\
&=\liminf_{n\to\infty}D(\Phi_n(P_n\rho P_n)\|\Phi_n(P_n\sigma P_n)),\label{eqnfirstininfiniteproof}
\end{align}
where the last equality follows from $\Phi_n(A)=\Phi_n(P_nAP_n)$ for all $A\in\cT(\cH)$.

Due to $\Phi_n(A)\in P'_n\cT(\cH')P'_n=\cT(P'_n\cH')$ for all $A\in\cT(\cH)$, one can view $\Phi_n$ as a mapping $\Phi_n:\cT(P_n\cH)\to\cT(P'_n\cH')$ between trace-class operators on finite-dimensional Hilbert spaces $P_n\cH$ and $P'_n\cH'$ when restricting its domain to $\cT(P_n\cH)=P_n\cT(\cH)P_n$. Using that the given linear map $\Phi$ is positive and trace-preserving, one easily verifies from (\ref{definePhin}) that $\Phi_n$ is a positive trace-preserving linear map on $\cT(P_n\cH)$. We can therefore apply our finite-dimensional result from Step 1 to (\ref{eqnfirstininfiniteproof}) and proceed:
\begin{equation}\nonumber
D(\Phi(\rho)\|\Phi(\sigma))\leq\liminf_{n\to\infty}D(P_n\rho P_n\|P_n\sigma P_n).
\end{equation}

We now apply the generalized Klein's inequality $D(A\|B)+\tr[B-A]\geq0$ for $A,B\in\cT_+(\cH)$ \cite{lanford-robinson,ohyapetz,wehrlreview} to $A:=P_n^\perp\rho P_n^\perp$ and $B:=P_n^\perp\sigma P_n^\perp$ with $P_n^\perp:={\mathbbm1}-P_n$ and continue:
\begin{align}
D(\Phi(\rho)\|\Phi(\sigma))&\leq\liminf_{n\to\infty} \Big\{D(P_n\rho P_n\|P_n\sigma P_n)+D(P_n^\perp\rho P_n^\perp\|P_n^\perp\sigma P_n^\perp)+\cdots\nonumber\\
&\qquad\qquad\qquad\qquad\qquad\qquad\cdots+{\rm tr}[P_n^\perp\sigma P_n^\perp-P_n^\perp\rho P_n^\perp]\,\Big\}\nonumber\\
&=\liminf_{n\to\infty}D(P_n\rho P_n+P_n^\perp\rho P_n^\perp\|P_n\sigma P_n+P_n^\perp\sigma P_n^\perp)\nonumber\\
&=\liminf_{n\to\infty}D(\Pi_n(\rho)\|\Pi_n(\sigma)),\label{eqnwithPi}
\end{align}
where we defined the completely positive and trace-preserving linear map $\Pi_n:\cT(\cH)\to\cT(\cH)$ by $\Pi_n(A):=P_nAP_n+P_n^\perp AP_n^\perp$ and used ${\rm tr}[P_n^\perp\rho P_n^\perp]\to 0$ and ${\rm tr}[P_n^\perp\sigma P_n^\perp]\to0$ as $n\to\infty$, and the fact that $P_n$ and $P_n^\perp$ project onto orthogonal subspaces.

At this point, we can conclude the desired statement $D(\Phi(\rho)\|\Phi(\sigma))\leq D(\rho\|\sigma)$ from (\ref{eqnwithPi}) by employing $D(\Pi_n(\rho)\|\Pi_n(\sigma))\leq D(\rho\|\sigma)$ for the completely positive trace-preserving maps $\Pi_n$, established by Lindblad in \cite{lindblad-cmp1975}; in fact, Lindblad's earlier monotonicity result for conditional expectations \cite{lindblad-cmp1974} is enough for this purpose.

To elucidate the minimal proof requirements, we would however like to remark that, for suitable choice of projections $P_n$, the last statement $D(\Pi_n(\rho)\|\Pi_n(\sigma))\leq D(\rho\|\sigma)$ can actually be deduced from much simpler facts than from the monotonicity results in \cite{lindblad-cmp1974,lindblad-cmp1975} which are based on Lieb's concavity theorem \cite{liebtheorem}. Namely, when the projections $P_n$ are all chosen to commute with $\rho$, so that $\Pi_n(\rho)=\rho$, the statement $D(\Pi_n(\rho)\|\Pi_n(\sigma))\leq D(\rho\|\sigma)$ becomes equivalent to the inequality ${\rm tr}[\rho\log\Pi_n(\sigma)]\geq{\rm tr}[\rho\log\sigma]={\rm tr}[\Pi_n(\rho)\log\sigma]={\rm tr}[\rho\Pi_n(\log\sigma)]$, which follows from the operator inequality $\log\Pi_n(\sigma)\geq\Pi_n(\log\sigma)$ that is due to operator concavity of the $\log$ function \cite{davis-schwarzineq}. Alternatively, when the projections $P_n$ are all chosen to commute with $\sigma$, the desired statement follows from the trace inequality ${\rm tr}[\Pi_n(\rho)\log\Pi_n(\rho)]\leq{\rm tr}[\rho\log\rho]$, i.e.\ the monotonicity of the von Neumann entropy under coarse-grainings, which is a classic result \cite{alberti-uhlmann-book,lindblad-cmp1974,ohyapetz,wehrlreview}.\qed % end of second proof (Theorem 1)

\bigskip
\bigskip

\noindent\textbf{Proof of Theorem \ref{theroremMainPrime}.}
Examining the proof of Theorem \ref{thm:Main}, we see that Step 1 requires only the trace-nonincreasing property along with the statement ${\rm tr}[\Phi(\rho)]={\rm tr}[\rho]$ instead of full trace-preservation.

For Step 2 to work, we need to ensure ${\rm tr}[\Phi_n(P_n\rho P_n)]={\rm tr}[P_n\rho P_n]$ for all $n$ even if $\Phi$ satisfies only the weaker conditions of Theorem \ref{theroremMainPrime} (from its definition, $\Phi_n$ is trace-nonincreasing and positive as $\Phi$ is). For this we choose the projections $P_n$ all to commute with $\rho$. In that case we have $\rho=P_n\rho P_n+P_n^\perp \rho P_n^\perp$ for $P_n^\perp:={\mathbbm1}_\cH-P_n$, and therefore
\begin{equation}
{\rm tr}[\Phi(P_n\rho P_n)]={\rm tr}[\Phi(\rho)-\Phi(P_n^\perp\rho P_n^\perp)]\geq{\rm tr}[\rho-P_n^\perp\rho P_n^\perp]={\rm tr}[P_n\rho P_n],
\end{equation}
where we used ${\rm tr}[\Phi(\rho)]={\rm tr}[\rho]$ and ${\rm tr}[\Phi(P_n^\perp\rho P_n^\perp)]\leq{\rm tr}[P_n^\perp\rho P_n^\perp]$. Together with the reverse inequality ${\rm tr}[\Phi(P_n\rho P_n)]\leq{\rm tr}[P_n\rho P_n]$, this finally gives the desired equality
\begin{equation}
{\rm tr}[P_n\rho P_n]={\rm tr}[\Phi(P_n\rho P_n)]={\rm tr}[\Phi_n(P_n\rho P_n)],
\end{equation}
where the last relation follows directly from the definition of $\Phi_n$.\qed % end of proof of Theorem 1'

\section{Conclusion}

We have shown that the quantum relative entropy $D$ \cite{umegaki1962} as well as the sandwiched Renyi divergences $D_\alpha$ for $\alpha\in(1,\infty)$ \cite{muller2013quantum,winter-wilde-yang} decrease monotonically under the application of any positive trace-preserving linear map. Our monotonicity proof for $D$ covers general separable underlying Hilbert spaces (Theorems \ref{thm:Main} and \ref{theroremMainPrime}), and the requirement of trace-preservation can be weakened (Theorems \ref{thm:DataProcSand} and \ref{theroremMainPrime}). Similar to some previous proofs of such monotonicity results for smaller classes of maps, the present proof uses complex interpolation arguments in an essential way; cf.\ the discussion in \cite{ruskai-review}. Our result constitutes the natural extension and minimal version of all previously known monotonicity results for the quantum relative entropy \cite{hiai2011,lindblad-cmp1975,ohyapetz,uhlmann-cmp1977}.

It remains to prove or disprove analogous monotonicity results for other divergence measures under general positive trace-preserving maps. The question is open in particular for the sandwiched Renyi-$\alpha$ divergences in the parameter range $\alpha\in(1/2,1)$ \cite{frank-lieb} (note that monotonicity holds for the case $\alpha=1/2$ \cite{milan-ogawa}, corresponding to the \emph{quantum fidelity} \cite{nielsenchuang}, whereas monotonicity is violated in general even for completely positive trace-preserving maps in the regime $\alpha\in[0,1/2)$ \cite{data-processing-violated}). For the ``old'' Renyi-$\alpha$ divergences defined by $\widetilde{D}_\alpha(\rho\|\sigma):=\log(\tr[\rho^\alpha\sigma^{1-\alpha}])/(\alpha-1)$, monotonicity under positive trace-preserving maps is known to hold for $\alpha\in\{0,2\}$, but seems to be an open question for $\alpha\in(0,2)\setminus\{1\}$ \cite{hiai2011}. The question is also open for other \emph{quasi-entropies} \cite{petz-monotone-metrics-see-conclusion}.

Our monotonicity results for positive maps motivate another set of interesting questions: Does equality in (\ref{equ:DataProcPos}) or (\ref{equ:DataProcPosAlpha}) for some positive trace-preserving linear map $\Phi$ and quantum states $\rho,\sigma$ imply the existence of a ``recovery map'', reversing the operation of $\Phi$ on these states? Such a reversibility statement for (\ref{equ:DataProcPos}) was first shown in seminal work by Petz \cite{petz-1988} for trace-preserving linear maps with stronger positivity properties such as complete positivity; see \cite{hiai2011} for a detailed investigation. Some equality conditions for the monotonicity relation (\ref{equ:DataProcPosAlpha}) of the sandwiched Renyi-$\alpha$ divergence have recently been established \cite{hiaimilan,jencova,nilanjana,milan-ogawa}. In particular, it was shown in \cite{jencova} that a recovery map exists if equality holds in (\ref{equ:DataProcPosAlpha}) for $\alpha=2$ and positive trace-preserving $\Phi$ or, alternatively, for some $\alpha\in(1,\infty)$ and $\Phi$ with somewhat stronger positivity properties.

More quantitatively, and in analogy to similar recent results for completely positive trace-preserving maps \cite{universal-recovery-map,goldenthompsen,sutter,wilde-recoverability}, one can ask whether recovery with high fidelity is possible if the decrease in relative entropy effected by a general positive trace-preserving map $\Phi$ in (\ref{equ:DataProcPos}) is small. This has been proven to be so for the special case of \emph{unital} trace-preserving positive maps and $\sigma$ the maximally mixed state \cite{buscemi-wilde}, giving also quantitative information about the well-known increase of the von Neumann entropy under such maps \cite{alberti-uhlmann-book,uhlmann-makov-master-eqns}. In the non-unital case, the known results rest crucially on complete positivity via the Stinespring dilation, or on the positivity of all tensor powers of $\Phi$, whereas just positivity is not yet known to be sufficient for this statement.

As described in the introduction, our results put the treatment of information processing within certain quantum-theoretical frameworks  \cite{weinberg-qm-vol2} onto more solid footing; on the other hand, they call into question several proposed definitions of non-Markovianity \cite{non-M-review}. These applications provide another strong motivation to further investigate the behaviour of general positive maps in contexts where mainly completely positive maps have been considered so far.

\subsection*{Acknowledgments}
We thank Francesco Buscemi, Milan Mosonyi, Volkher Scholz, Daniel Stilck Fran\c{c}a, and Mark Wilde for useful comments. AMH acknowledges financial support from the European Research Council (ERC Grant Agreement no 337603), the Danish Council for Independent Research (Sapere Aude) and VILLUM FONDEN via the QMATH Centre of Excellence (Grant No.\ 10059). DR acknowledges funding from the ERC grant DQSIM.

\end{document}